\documentclass[pra,notitlepage,floatfix,superscriptaddress]{revtex4-1}
\usepackage{dcolumn,amsmath}
\usepackage{graphicx}
\usepackage{bm}
\usepackage{amssymb}
\usepackage{multirow}

\usepackage{amsfonts}
\usepackage{amsmath}
\usepackage{cancel}
\usepackage{xcolor}
\begin{document}
\title{Symmetric top molecule  YbOCH$_3$ in the fundamental $\mathcal{P}$, $\mathcal{T}$-violation searches}

\author{Anna Zakharova} \email{zakharova.annet@gmail.com} 
%
\affiliation{St. Petersburg State University, St. Petersburg, 7/9 Universitetskaya nab., 199034, Russia} 
\affiliation{Petersburg Nuclear Physics Institute named by B.P. Konstantinov of National Research Centre
"Kurchatov Institute", Gatchina, 1, mkr. Orlova roshcha, 188300, Russia}
\date{Received: date / Revised version: date}
\begin{abstract}{
The symmetric top molecule YbOCH$_3$ is studied for its potential to $\mathcal{P}$, $\mathcal{T}$-violation searches.
The influence of the rotations and vibrations of the YbOCH$_3$ on such violating effects as the electron electric dipole moment (eEDM) and the scalar-pseudoscalar electron-nucleon interaction (Ne-SPS) is studied using the coupled channels method. The corresponding sensitivity parameters $E_{\rm eff}$ and $E_{\rm s}$ are computed.
} 
\end{abstract}

\maketitle
\section{Introduction}
\label{intro}
Studying of discrete symmetry violations such as space reflection ($\mathcal{P}$), charge conjugation ($\mathcal{C}$), and time reverse ($\mathcal{T}$) is a powerful way in New physics searches \cite{khriplovich2012cp}. 
The modern theory of fundamental interactions - the Standard model (SM) \cite{schwartz2014quantum,particle2020review},  includes the $\mathcal{CP}$ ($\mathcal{T}$) symmetry violation due to the complex-valued Cabibbo-Kobayashi-Maskawa (CKM) \cite{Cabibbo1963,KobayashiMaskawa1973} and Pontecorvo–Maki–Nakagawa–Sakata (PMNS)  \cite{Pontecorvo1957,MNS1962} mixing matrices.
But some effects, associated with the discrete symmetries violation in SM \cite{schwartz2014quantum,particle2020review}, are strongly suppressed. The examples are the electron electric dipole moment (eEDM) and scalar-pseudoscalar electron-nucleon interaction (S-PS) \cite{Fukuyama2012,PospelovRitz2014,YamaguchiYamanaka2020,YamaguchiYamanaka2021}. This opens a way to experimentally test many proposals for the new physics beyond the Standard Model that predict larger values for eEDM and S-PS coupling constant.

The limit on $\mathcal{P}$, $\mathcal{T}$-odd scalar-pseudoscalar electron-nucleon interaction can be put in molecular experiments \cite{ginges2004violations,PospelovRitz2014,ChubukovLabzowsky2016}. The eEDM searches can be performed in the same molecular experiments \cite{baron2014order,ACME:18}.
 The enhancement parameters for such $\mathcal{P}$, $\mathcal{T}$-violating effect can be obtained in quantum chemical calculations \cite{KozlovLabzowsky1995, titov2006d, Safronova2017}.
The best EDM limits were received in diatomic molecules HfF$^{+}$ \cite{Cornell:2017,Petrov:18,roussy2023improved} and ThO\cite{ACME:18,DeMille:2001,Petrov:14,Vutha:2010,Petrov:15,Petrov:17}. The most resent upper bond limit was obtained in \cite{roussy2023improved} with HfF$^{+}$ ions. The advantage of such molecules is the presence of the levels of opposite parity in their spectrum, that allow techniques to reduce systematic effects. While the diatomics have only electronic parity doublets the polyatomic molecules also admit the rovibrational states such as $l$-doublets and, in case of the symmetric top molecules, $K$-doublets.

Some triatomic molecules with parity doublets can be cooled by lasers \cite{kozyryev2017sisyphus,steimle2019field,augenbraun2020laser}. The influence of vibrations and rotations on sensitivity $E_{\rm eff}$ and $E_{\rm s}$ to eEDM and S-PS interactions were earlier studied for triatomic molecules \cite{gaul2020ab,ourRaOH,zakharova2021rovibrational}.

Symmetric top molecules with general formula (M)OCH$_3$, such as RaOCH$_3$ \cite{zakharova2022rotating}, and YbOCH$_3$ also allow lazer-cooling \cite{isaev2016polyatomic, kozyryev2016proposal,kozyryev2019determination,augenbraun2021observation} and can have non-zero value of the projection of $K$ in molecule-fixed frame in the ground state, that facilitates the experimental study of symmetrical top type molecules. Laser cooling of another symmetric top, CaOCH$_3$, demonstrate the experimental progress and perspective in studying of such molecules \cite{mitra2020direct}. First controlled syntheses of cold symmetric-top ion  RaOCH$_3^+$ was achieved in \cite{fan2021optical}. 

 Theoretical considerations are also in progress. Earlier, there were studied the $^{225}$RaOCH$_3^+$ for Schiff moment search in \cite{yu2021probing}. Analytic-gradient-based method for relativistic coupled-cluster calculations for eEDM was developed, and applied to the symmetric top molecules \cite{zhang2021calculations}. The enhancement parameters for optimised geometry of BaCH$_3$ and YbCH$_3$ symmetric top were discussed in \cite{chamorro2022molecular}
It was shown that enhancement parameters of  the parity-violating effects in closed-shell radium-containing molecule such as RaOCH$_3^+$ expected to be large, and open up great prospects for experimentional search \cite{gaul2023mathcal}.

However, this work was done for the optimized geometries, neglecting the impact of the rotations and vibrations of such molecules. In contrast, as was noted above, the actual experiments are to be performed on the rotational and vibrational parity doublets. Recent succesful experiment with optical trapping of a polyatomic molecule CaOH in an $l$-doublet state makes the theoretical study of parity violation effects on parity doublets highly relevant \cite{hallas2023optical}

While this problem was extensively studied for the triatomic molecules \cite{gaul2020ab,ourRaOH,zakharova2021rovibrational,zakharova2022impact}, for the symmetric top molecules this was only done in the harmonic approximation for RaOCH$_3$ molecule in \cite{zakharova2022rotating}. While the harmonic approximation may be a first good estimate it does not allow to easily study the impact and interplay of all rotational and vibrational effects and of the potential anharmonicities. In this paper we apply the coupled channels technique earlier succesfully used to investigate the rovibrational dynamics of the triatomic molecules \cite{zakharova2021rovibrational,petrov2022sensitivity,zakharova2022impact} to the symmetric top molecule YbOCH$_3$.

\section{The geometry of the molecule}
\label{Sec:Geometry}
\begin{figure}[h]
\centering
  \includegraphics[width=0.5\textwidth]{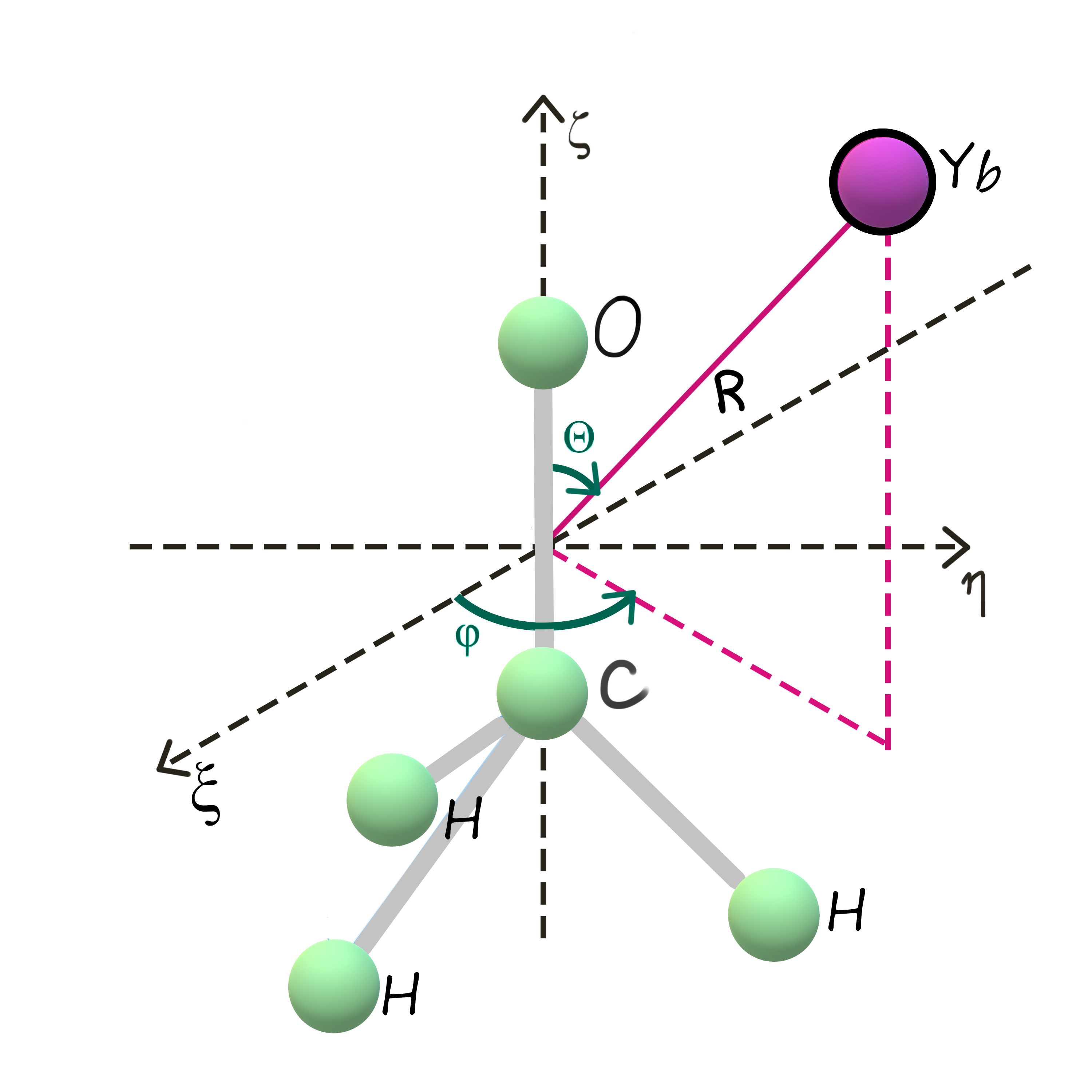}
  \caption{The YbOCH$_3$ molecule}
  \label{fig:molecule}
\end{figure}

In the Born-Oppenheimer approximation the molecular wavefunction is represented as a product of two parts: electronic and nuclear,\begin{equation}
\Psi_{\rm total}\simeq \Psi_{\rm nuc}(R,\hat{R},\hat{r},\gamma)\psi_{\rm elec}(\{\vec{r}_i\}|R,\theta,\varphi),
\label{psiexp}
\end{equation}
where $R$, $\theta$ and $\varphi$ define the geometry of the YbOCH$_3$ molecule Fig.~\ref{fig:molecule}.The angle $\gamma$ is orientation of the CH$_3$ around $\zeta$. The unit vectors along Yb -- ligand c.m. axis,$\hat{R}$, and ligand $\zeta$ axis,$\hat{r}$, also determine the configuration. 

We consider the rigid ligand since the Yb -- OCH$_3$ bond stretching and bending frequencies  are much smaller than the vibrations of the OCH$_3$. The geometry of the ligand is similar to the one considered in \cite{zakharova2022rotating}. The geometric parameters of the molecule are presented in the Table \ref{tbl:LigandGeom}.

\begin{table}[h]
\small
  \caption{The ligand geometry}
  \label{tbl:LigandGeom}
  \renewcommand{\arraystretch}{1.5}
  \begin{tabular*}{0.48\textwidth}{@{\extracolsep{\fill}}ll}
    \hline\hline
$r(O-C)$ & $2.600\,\mathrm{a.u.}$\\
$r(C-H)$ & $2.053\,\mathrm{a.u.}$ \\
$\angle(O-C-H)$ & $110.73^\circ$ \\
\hline\hline
  \end{tabular*}
\end{table}

The effective Hamiltonians for $\mathcal{P}$, $\mathcal{T}$-odd interactions considered are, 
\begin{align}
&\hat{H}_{\rm d}=  2d_e\sum_{i}
  \left(\begin{array}{cc}
  0 & 0 \\
  0 & \bf{\sigma_i E_i} \\
  \end{array}\right)\ 
 \label{Hd},\\
&\hat{H}_{\rm s}=ik_s\frac{G_F}{\sqrt2}\sum_{j=1}^{N_{elec}}\sum_{I=1}^{N_{nuc}}{\rho_I\left(\vec{r_j}\right)Z_I}\gamma^0\gamma^5.
\end{align}
Here $\bf{E_i}$ is the inner molecular electric field acting on i-th electron, $\rho_I$ is the charge density of the $I$-th nucleon normalized to unity. $\hat{H}_{\rm d}$ corresponds to interaction of the electron electic dipole moment with electronic shell of the YbOCH$_3$ molecule. $\hat{H}_{\rm s}$ corresponds 
to the scalar-pseudoscalar interaction between the electrons and the nuclei.

The perturbation by these Hamiltonians is described by the enhancement parameters $E_{\rm eff}$ and $E_{\rm d}$,
\begin{align}
E_{\rm eff}(R,\theta,\varphi)=\frac{\langle\psi_{elec}(R,\theta,\varphi)| \hat{H}_{\rm d}|\psi_{elec}(R,\theta,\varphi)\rangle}{d_e{\rm sign}(\Omega)},\\ E_s(R,\theta,\varphi)=\frac{\langle\psi_{elec}(R,\theta,\varphi)| \hat{H}_{\rm s}|\psi_{elec}(R,\theta,\varphi)\rangle}{k_s{\rm sign}(\Omega)}.
\end{align}

After electronic calculation of these parameters for the fixed geometries we average  them over the rovibrational nuclear wavefunction $\Psi_{\rm nuc}$ to take into account rotations and vibration of the molecule.

\section{Coupled-channel technique}
\label{Sec:СС}
Assuming that OCH$_3$ ligand is rigid, we divide the YbOCH$_3$ molecule into two systems. The first one is a linear rotor with length $R$ formed by the heavy atom Yb and the ligand center of mass, to which we attribute angular momentum $\hat{\vec{l}}$. The associated quantum numbers are: $l$ for $\hat{\vec{l}}^2$ and $m_l$ for its projection on the $z$ axis of the space-fixed frame (SFF).
The second one corresponds to ligand OCH$_3$, rigid symmetric top, with the angular momentum $\hat{\vec{j}}$. The associated quantum numbers are: $j$ for $\hat{\vec{j}}^2$, $m$ for its projection on the $z$ axis, and $k$ for its projection on the $\zeta$ axis of the ligand-fixed frame (LFF).

The adiabatic Hamiltonian of nuclear motion can be written as:
\begin{equation}
    \hat{H}_{nuc }=\left[-\frac{1}{2 \mu} \frac{\partial^2}{\partial R^2}+\frac{\hat{\vec{l}}^2}{2 \mu R^2}+V \left( R, \hat{R}, \hat{r}, \gamma\right)\right]+\hat{H}_{lig},
\end{equation}
where the first term is the kinetic energy of longitudinal deformations, the second corresponds to the kinetic energy of the linear rotor Yb-ligand rotation. The rigid ligand Hamiltonian is
\begin{equation}
\hat{H}_{\text{lig}}=\frac{\hat{j}_{\xi}}{2 I_{\xi}}+\frac{\hat{j}_{\eta}}{2 I_{\eta}}+\frac{\hat{j}_{\zeta}^2}{2 I_{\zeta}},
\end{equation}
where $\xi,\eta, \zeta $ - axis in the coordinate system associated with the ligand.

The ligand we are considering is a symmetric top i.e. $I_\xi=I_\eta > I_\zeta$. Its Hamiltonian then can be rewritten as,
\begin{equation}
\hat{H}_{\text{lig}}=B\hat{j}^2+(A-B)\hat{j}_z^2,
\end{equation}
where the rotational constants $A$ and $B$ can be found through the ligand moments of inertia,
\begin{equation}
B=\frac{1}{2I_{\xi}},\quad A=\frac{1}{2I_{\zeta}}.
\end{equation}
The ligand Hamiltonian eigenstates have definite $j$, $m$ and $k$ eigenvalues and the corresponding energy equals to,
\begin{equation}
E_{\rm lig}^{jk}=Bj(j+1)+(A-B)k^2.
\end{equation}

We are looking for a solution of the Schrödinger equation in the form,
\begin{equation}
\hat{H}_{\text{nuc}} \psi^{JM}(R,\hat{R},\hat{r},\gamma)=E \psi^{JM}(R,\hat{R},\hat{r},\gamma),
\end{equation}
where the unit vectors $\hat{R}, \hat{r}$ and angle $\gamma$ - specify the orientation of the molecule in the space-fixed frame.

As a way to discretize the dependence of $\psi^{JM}$ on the continuous angles let us decompose into the eigenfunctions of the angular momenta $\hat{\vec{j}}^2, \hat{j}_\zeta$ and $\hat{\vec{l}}^2$:
\begin{equation}
\psi^{J M}(R, \hat{R}, \hat{r}, \gamma)=\sum_{j=0}^{\infty} \sum_{k=-j}^{+j} \sum_{l=0}^{+\infty} F_{j k l}^J(R){Y}_{j k l}^{JM}(\hat{R}, \hat{r}, \gamma).
\end{equation}

These eigenfunctions can be represented as,
\begin{equation}
Y_{jkl}^{J M}(\hat{R}, \hat{r}, \gamma)=\sum_{m=-j}^{+j} \sum_{m_l=-l}^{l} \langle j m l m_l|JM\rangle Y_{l m_l}(\hat{R}) D_{j m k}(\hat{r}, \gamma),
\end{equation}

where $\langle j m l m_l|JM\rangle$ - Clebsh-Gordan coefficients;
$Y_{l m_l}(\hat{R})$ - spherical harmonic, which is an eigenfunction of the Hamiltonian of a free rigid linear rotor; $D_{j m k}(\hat{r}, \gamma)$ - Wigner function, the eigenfunction of the Hamiltonian of a free rigid symmetric top.

After this expansion the equation for the nuclear wave function becomes,

\begin{equation}
    \Big(\frac{d^2}{d R^2} -\frac{l(l+1)}{ R^2} +2\mu E - 2\mu E_{\rm lig}^{jk}\Big)F_{j k l}^J(R) \\
    \\= 2 \mu \sum_{j'k'l'} v_{jklj'k'l'}(R)F_{j' k' l'}^J(R),
\end{equation}

which may be considered a one-dimensional Schr\"{o}dinger equation for the multi-component wavefunction with the matrix potential.
The
elements of this matrix potential are expressed through 3j and 6j symbols:

\begin{multline}
v_{jkl,\tilde{j}\tilde{k}\tilde{l}}(R)=(-1)^{j+\tilde{j}+\tilde{k}-J}\sqrt{\frac{(2j+1)(2\tilde{j}+1)(2l+1)(2\tilde{l}+1)(2\lambda+1)}{4\pi}}\\
\times\sum_{\lambda\mu} V_{\lambda\mu}(R)
\begin{Bmatrix}j&l&J\\\tilde{l}&\tilde{j}&\lambda\end{Bmatrix}\begin{pmatrix}l&\lambda&\tilde{l}\\0&0&0\end{pmatrix}\begin{pmatrix}j&\lambda&\tilde{j}\\-k&\mu&\tilde{k}\end{pmatrix},
\end{multline}
with $V_{\lambda\mu}(R)$ obtained by expanding the adiabatic potential,
\begin{equation}
V(R, \hat{R}, \hat{r})=V(R, \theta, \varphi)=\sum_{\lambda=0}^{+\infty} \sum_{\mu=-\lambda}^{+\lambda} V_{\lambda \mu}(R) Y_{\lambda \mu}( \theta, \varphi).
\end{equation}

As result, we turn the Schr\"{o}dinger equation on $\Psi^{JM}$ into an infinite system of equations on a discrete set of unknown functions $F_{j k l}^{JM}(R)$ that depend only on $R$. As high values of angular momenta quantum numbers should be irrelevant for the low energy behaviour we can cut off the sum over the quantum numbers,
\begin{equation}
\lambda\leq\lambda_{\rm max},\quad j\leq j_{\rm max},\quad l\leq l_{\rm max}.
\end{equation}
This makes a system of equations finite and tractable with the numerical methods. To deal with the dependence on $R$ we introduce a basis of functions $\Phi_a(R)$:
\begin{align}
& F(R) \mapsto F_{a}=\int_0^{+\infty} dR\,\Phi_{a}^\ast(R) F(R), \\
& \hat{O} \mapsto O_{ab} =\int_0^{+\infty} d R \,\Phi_{a}^\ast(R) \hat{H}\Phi_{b}(R).
\end{align}
Restricting this basis to be finite we turn the system of the differential equations into a finite algebraic system.

The described approach makes it possible to take into account the anharmonicites of the potential and all the rotational and vibrational effects. It also may be easily generalized to consider the non-adiabatic transitions and extra interactions if one may compute the corresponding matrix elements.

\section{Application to YbOCH$_3$}

In this work we will approximate the potential with,
\begin{equation}
V(R,\theta,\phi)\simeq V_R(R)+V_\theta(\theta),\quad V_\theta(0)=0
\end{equation},
where the longitudinal part is approximated by the harmonic potential,
\begin{equation}
V_R(R)=\mu\frac{\omega_R^2}{2}(R-R_{\rm eq})^2,
\end{equation}
and for the basis of the functions of $R$ we choose the eigenbasis for this potential,
\begin{align}
\Big(\frac{d^2}{dR^2}-2\mu V_R(R)\Big)\Phi_a(R)=-2\mu\omega_R \Big(a+\frac{1}{2}\Big)\Phi_a(R),\\
\Phi_a(R)=\frac{1}{2^a a!}\Big(\frac{\mu\omega_R}{\pi}\Big)^\frac{1}{4}e^{-\frac{m\omega_R}{2}(R-R_{\rm eq})^2}H_a\Big(\sqrt{m\omega_R}(R-R_{\rm eq})\Big),
\end{align}
where $H_a$ is the Hermite polynomial. Consider the only remaining term non-trivially affecting $R$-dependence that corresponds to the centrifugal potential. Near the equilibrium it can be approximated as,
\begin{equation}
\frac{l(l+1)}{R^2}\simeq \frac{l(l+1)}{R_{\rm eq}^2}\Big(1-2\frac{R-R_{\rm eq}}{R_{\rm eq}}\Big).
\end{equation}
In this approximation, for definite $l$ it modifies the harmonic potential simply by moving the equilibrium point and shifting the overall energy. I.e. it results in a centrifugal stretching of the Yb -- ligand bond.

Note that for sufficiently small shifts of the equilibrium point the ground state of the oscillator may be represented as,
\begin{equation}
\Phi_0(R-\delta R)\simeq\Phi_0(R)+\sqrt{\frac{\mu\omega_R}{2}}\delta R\Phi_1(R).
\end{equation}
So, while it may seem a very rough approximation taking just $\Phi_0$ and $\Phi_1$ as basis functions may be enough to take into account a centrifugal stretching to a certain degree. Then the centrifugal term restricted to this basis may be represented as a matrix,
\begin{equation}
\frac{l(l+1)}{R^2}\Big\vert_{\{\Phi_0(R),\Phi_1(R)\}} \simeq \frac{l(l+1)}{R_{\rm eq}^2} \begin{pmatrix}1&-\sqrt{\frac{2}{m\omega R_{\rm eq}^2}}\\-\sqrt{\frac{2}{m\omega R_{\rm eq}^2}}&1\end{pmatrix}.
\end{equation}

\section{Electronic computations}
\label{Sec:electronic}

The molecular orbitals were obtained in DIRAC19 software package at Dirac-Harthree-Fock self-consistent field (SCF) level. To reduce the computational cost for the molecule with heavy Ytterbium atom we employed  28-electron basis  with a generalized relativistic effective core potential (GRECP) with spin-orbit interaction blocks \cite{titov1999generalized,mosyagin2010shape,mosyagin2016generalized}, developed by the Quantum Physics and Chemistry Department of the PNPI \cite{QCPNPI:Basis}. Previously, this basis was used in calculations of the eEDM and S-PS elecron-nucleon interaction  enhancement parameters of the YbOH molecule \cite{zakharova2021rovibrational,zakharova2022impact}.

For the correlation computations we use the relativistic coupled cluster method (CCSD) realized in DIRAC19 software package. Based on the gaps in the molecular orbital spectrum, we restrict the active space to the 18 occupied orbitals balanced by 18 virtual orbitals.

As the molecular spinors, calculated with GRECP have incorrect behavior in the core region, we construct the four-component spinors with help of the one-center restoration procedure \cite{Petrov:02,titov2006d}. This technique is implemented in the MOLGEP program  that allows us to obtain the matrix elements of $\mathcal{P}$, $\mathcal{T}$- violating parameters on the corrected molecular spinors. 

We use the finite field method to obtain the enhancement parameters of scalar-pseudoscalar electron-nucleon interaction and electron EDM at the CCSD level.
This approach consists of adding a small perturbation of the property $\hat{O}$ under consideration  to the Hamiltonian:
\begin{equation}
    \hat{H}(\epsilon)\equiv\hat{H}+\epsilon\hat{O}.
\end{equation}
Because the Kramers-restricted nature of the DIRAC SCF computation and our use of the matrix elements computed in MOLGEP, we introduce such perturbation as a modification of the one-electron integrals \cite{zakharova2022rotating}
We carry out further CCSD calculations in the DIRAC program with modified integrals.

Energy level of stationary state $|\psi\rangle$ shifts this way
\begin{equation}
    E(\epsilon) = E + \epsilon\langle\psi| \hat{O}|\psi\rangle+O(\epsilon^2).
\end{equation}

The expectation values of properties $\hat{O}$ are extracted from the CCSD energies obtained for different $\epsilon$ values
\begin{equation}
    \langle\psi|\hat{O}|\psi\rangle\simeq \frac{E(+\epsilon)-E(-\epsilon)}{2\epsilon}.
\end{equation}

\section{Results and discussion}
\label{Sec:results}

The potential surface at CCSD level results in the equilibrium distance $R_{\rm eq}=5.25\, \mathrm{a.u.}$ and in the vibrational frequencies $\omega_\perp = 173.7664 \,\rm{cm}^{-1}$, $\omega_\parallel = 349.2547\, \rm{cm}^{-1}$. This may be compared with the experimental values of $\omega_\perp = 130\pm 5\, \rm{cm}^{-1}$, $\omega_\parallel = 400\pm 5\, \rm{cm}^{-1}$ \cite{augenbraun2021observation}.

\begin{figure}[h]
\centering
\includegraphics[width=0.5\textwidth]{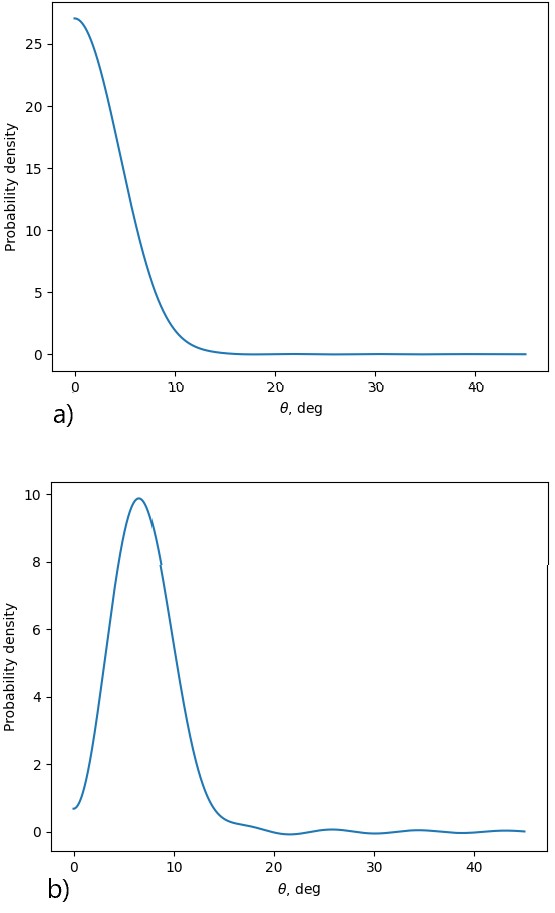}
  \caption{The $\phi$-averaged probability densities for the rovibrational wavefunctions obtained using the coupled channel technique (a) $(J=0, v_\perp=0, K=0)$ ground state (b) $J=1, v_\perp=1, l=\pm 1, K=\pm 1)$ first $l$-doublet}
  \label{fig:wf1}
\end{figure}

The properties were computed for the equilibrium $R=R_{\rm eq}$ and for the three values of the azimuthal angle $\phi =0, 30^\circ, 60^\circ$. The properties are interpolated in a way consistent with the symmetries of the molecule,
\begin{equation}
E_{\rm eff, s}=E_{\rm eff, s}^{(0)}(\theta) + E_{\rm eff, s}^{(1)}(\theta)\cos 3\phi + E_{\rm eff, s}^{(2)}(\theta)\cos 6\phi.
\end{equation}
The rovibrational wavefunctions were found for $j_{\rm max}=l_{\rm max}=30$ and $k_{\rm max}=28$. When $\phi$-dependence of the potential is neglected, the expantion contains only $\mu$ = 0 terms. Then, due to the properties of the 3j-symbol, the potential couples only states with $k$  = $k'$. Therefore, $k$ is a good quantum number for the resulting rovibrational wavefunctions. Then, as long as $k<3$ only $\phi$-averaged values of the properties $E_{\rm eff, s}^{(0)}(\theta)$ contribute to the resulting property expectation value. Then one may represent it as,
\begin{equation}
\langle E_{\rm eff,s}\rangle=\int dR \,d\hat{R}\, d\hat{r} \,d\gamma \Big|\Psi_{\rm nuc}\Big|^2 E_{\rm eff,s}(\theta,\phi)=\int_0^\pi d\theta \sin\theta\, \rho_\Psi(\theta) E_{\rm eff,s}^{(0)}(\theta),
\end{equation}
where $\rho_\Psi$ is a probability distribution corresponding to $\Psi_{\rm nuc}$ rovibrational state averaged over rotational degrees of freedom, $R$ and $\phi$. Such distributions are represented on Fig. \ref{fig:wf1} 
Their shape is close to the one expected from the harmonic approximation considerations.

The values of the enhancement parameters averaged over the rovibrational wavefunctions for several low-lying states are given in Table \ref{tbl:table1}.

\begin{table}[h]
\small
  \caption{The $\mathcal{P}$, $\mathcal{T}$-odd parameters for the rovibrational states of the YbOCH$_3$ molecule}
  \label{tbl:table1}
  \renewcommand{\arraystretch}{1.5}
  \begin{tabular*}{0.48\textwidth}{@{\extracolsep{\fill}}lllllll}
    \hline\hline
$v_{\parallel}$ & $v_\perp$ &  $l$ & $K$ &  $(E-E_0)/\omega_\perp$ & $E_{\rm eff},\, \frac{\rm GV}{\rm cm}$ &$E_s,\, {\rm kHz}$\\
\hline
  \multicolumn{1}{l}{J=0}\\
\hline
0 & 0 & 0 & 0 & 0   & 24.693    & 21.423\\
0 & 1 & $\pm 1$ & 0 &1.028 & 24.666  & 21.504\\
1 & 0 & 0 & 0 &2.009 & 24.551  & 21.325\\
0 & 2 & 0 & 0 &2.013 & 23.880  & 20.866\\
0 & 2 & $\pm 2$ & 0 &2.117 & 24.135  & 21.129\\
\hline
  \multicolumn{1}{l}{J=1}\\
  \hline
0 & 0 & 0 & 0 & 0.001   & 24.693    & 21.423\\
0 & 0 & 0 & $\pm 1$ &0.032 & 24.693  & 21.423\\
0 & 1 & $\pm 1$ & $\pm 1$ &1.002 & 24.666  & 21.504\\
0 & 1 & $\pm 1$ & 0 &1.029 & 24.666  & 21.504\\
0 & 1 & $\pm 1$ & $\mp 1$ &1.117 & 24.665  & 21.504\\

 \hline\hline
  \end{tabular*}
\end{table}

A more accurate description of the potential and inclusion of the spin-orbit and spin-spin interactions would allow us to extract the values of $l$-doubling or even $K$-doubling from the energies of these states. We also note that for $k\geq 3$ we may expect the contribution from $E_{\rm eff,s}^{(1)}$ term that may lead to the nontrivial parity-violating mixing between $K$-doubled states. However, these questions lie beyond the scope of the current paper and we leave them for the future work.

A comparison of the values of the enhancement parameters obtained for a six-atomic molecule of the symmetrical top type with other molecules is presented in the table

\begin{table}[h]
\small
  \caption{The comparison
of the $\mathcal{P}$, $\mathcal{T}$-odd parameters for different molecules}
  \label{tbl:tableResults}
  \renewcommand{\arraystretch}{1.5}
  \begin{tabular*}{0.48\textwidth}{@{\extracolsep{\fill}}lllll}
    \hline\hline
& $v_\perp$ &  $K$ & $E_{\rm eff},\, \frac{\rm GV}{\rm cm}$ &$E_s,\, {\rm kHz}$\\
\hline
YbOCH$_3$ & \multicolumn{2}{l}{equilibrium}   &   24.943  & 21.529\\
YbOCH$_3$ & $v_\perp=0$ & $0$   &   24.693  & 21.423\\
& & $\pm 1$   &   24.693  & 21.423\\
YbOCH$_3$ & $v_\perp=1$, $l=\pm1$\textbf{} & $0$   &   24.666  & 21.504\\
& & $\pm 1$   &   24.666  & 21.504\\
\hline
\hline
RaOCH$_3$\cite{zakharova2022rotating} & \multicolumn{2}{l}{equilibrium}   &   48.346  & 64.015\\
RaOCH$_3$\cite{zakharova2022rotating} & $v_\perp=0$ & $0$   &   47.930  & 63.436\\
& & $1$   &   47.929  & 63.435\\
RaOCH$_3$\cite{zakharova2022rotating} & $v_\perp=1$, $l=\pm1$ & $0$   &   47.649  & 63.064\\
& & $\pm 1$   &   47.648  & 63.063\\
\hline
YbOCH$_3$\cite{zhang2021calculations} & \multicolumn{2}{l}{equilibrium}  & 24.0 & --\\
    \hline
RaOH\cite{ourRaOH} & \multicolumn{2}{l}{equilibrium}  & 48.866  & 64.788\\
& $v=1$ &      & 48.585 & 64.416\\
\hline
YbOH\cite{zakharova2021rovibrational} & \multicolumn{2}{l}{equilibrium} & 23.875 & 20.659 \\
& $v=1$ &
  & 23.576 & 20.548 \\

 \hline\hline
  \end{tabular*}
\end{table}


\section{acknowledgement}
The work was supported by the Theoretical Physics and Mathematics Advancement Foundation “BASIS” (grant N\textsuperscript{\underline{o}} 23-1-4-10-1).


\end{document}